\documentstyle[11pt]{article}

\newcommand{\bcn}{\begin{center}}
\newcommand{\beq}{\begin{equation}}
\newcommand{\beqn}{\begin{eqnarray}}
\newcommand{\ecn}{\end{center}}
\newcommand{\eeq}{\end{equation}}
\newcommand{\eeqn}{\end{eqnarray}}

 \def\lsim{\mathrel{\rlap{\lower4pt\hbox{\hskip1pt$\sim$}}
    \raise1pt\hbox{$<$}}}
\def\slash#1{\setbox0=\hbox{$#1$}#1\hskip-\wd0\hbox to\wd0{\hss\sl/\/\hss}}
\begin{document}

\rightline{KU-HEP-93-26}
\vspace*{2.0cm}
\bcn
{\large\bf Evidence for Observation of Color
Transparency in\\ Proton Nucleus Collisions
\footnote{To appear in the proceedings of the Workshop on
Future Directions in Particle and Nuclear Physics at Multi-GeV Hadron
Facilities, BNL, March 1993, edited by D. Geesaman et al.}}\vspace*{1.0cm}

{\bf Pankaj Jain and John P. Ralston} \\

{\it Department of Physics and Astronomy \\
The University of Kansas \\
Lawrence, KS-66045-2151\\ }\vspace*{1.0cm}

\vspace*{3.0cm}
{\bf ABSTRACT}
\ecn
\noindent
We introduce a data analysis procedure for color transparency
experiments which is  considerably less model dependent than the transparency
ratio method. The new method is based  on fitting the shape of the A dependence
of the nuclear cross section at fixed momentum transfer to  determine the
effective attenuation cross section for hadrons propagating through the
nucleus. The hard scattering cross section is then determined directly
from the data.
We apply this procedure to the Brookhaven experiment of
Carroll et al and find  that it clearly shows color transparency: the effective
attenuation cross section in events with  momentum transfer $Q^2$ is
approximately
$40\ mb\ (2.2\ GeV^2/Q^2)$.  The fit to the data also supports  the idea that
the hard
scattering inside the nuclear medium is closer to perturbative QCD predictions
than is the scattering of isolated protons in free space. We discuss
the application of our approach to electroproduction experiments.

\vspace*{1.2cm}
\hspace*{3pc}
\vfill
\eject

\noindent {\bf 1.} Color transparency [1] is a theoretical prediction that
under certain
circumstances the strong  interactions may appear to be effectively reduced in
magnitude. Consider an exclusive
 quasi elastic process in which an incoming proton
knocks a proton out of a nucleus with large momentum transfer
without disturbing the rest of the nucleus.
If only short distance components of the
proton wave function contribute to this process,
then the incoming and outgoing protons will have considerably reduced
attenuation as they propagate through the nucleus.
However, to  make a
quantitative measurement of the attenuation, one must have a value for the hard
scattering  rate.  In the absence of a normalization of the hard sub-process,
 only a combination of  the hard scattering rate and
attenuation rate is measured in an experiment.

\medskip

Theory at present cannot supply absolute normalizations for exclusive
processes, so the scattering  rate in an isolated hadron has been used
previously as a benchmark for comparison with the nuclear  target. The
"transparency ratio" $T(Q^2,\ A)$
was introduced by Carroll et al [2]:  it is the
ratio of a  cross section measured in the nuclear target to the analogous cross
section for isolated hadrons in  free space. However it has become clear
that this ratio is not directly related to the attenuation
in the nuclear medium. Any soft components of the hadronic wave function
 contributing to the free space cross section should be
filtered by the nucleus, resulting in the survival of a short
distance part of the wave function [3]. Therefore the free space cross
section may be quite different from the hard scattering cross section
inside the nuclear target. If only the hard components contribute
to the nuclear hard scattering, we expect that it will be closer
to the perturbative QCD (pQCD) predictions. In contrast, the free space cross
section is known to deviate considerably from pQCD.

\medskip

In order to avoid making too many assumptions about the hard scattering
we introduce a more systematic data analysis procedure which
does not invoke the free space scattering. For definiteness
we concentrate on the $pA\rightarrow p'p''(A-1)$ process,
although the procedure is similar for
electroproduction experiments $eA\rightarrow e'p(A-1)$.
The procedure
is based on fitting the A dependence of the nuclear cross section
$d\sigma_A$ at fixed energy to determine the effective attenuation
cross section $\sigma_{eff}$. Assuming that the hard scattering
factorizes from the subsequent propagation of the hadron,
$d\sigma_A$ can be written as

\beq
{d\sigma_{_A}/dt\over Z}=s^{-10}N(Q^2)\; P_{ppp}(\sigma_{eff}(Q^2),\; A)\ \ ,
\eeq
where $s^{-10}N(Q^2)$ is the hard scattering cross section, $P_{ppp}$
is the survival probability and $Q$ is the characteristic momentum
scale in the collision, $Q^2\approx -t$.
The factors of $s^{-10}$ and 1/Z are simply definitions to take out some
typical
orders of magnitude -  their usage will introduce no bias.
By fitting the $A$ dependence of $d\sigma_A$ at fixed $Q^2$ we can
determine the functional dependence of $P_{ppp}$ on $A$. In order to
extract an effective cross section $\sigma_{eff}$ at a given energy
we model the survival probability
by assuming exponential attenuation.
The survival factor is then obtained by using a
Monte-Carlo event generator which propagates the incoming proton to the
hard scattering point, and follows the outgoing protons along
kinematically allowed paths. The hard scattering is assumed to
occur at random over the volume of the nucleus.

By following this procedure we can determine $\sigma_{eff}(Q^2)$ and
$N(Q^2)$ at each energy. Since $\sigma_{eff}(Q^2)$ and
the normalization $N(Q^2$) are free, the fit to  the data may or may not show
that these parameters vary with energy. The best fit for the two
energies reported by Carroll et al. [2] is shown in fig. (1a) and (1b).

\medskip
\noindent The resulting $\sigma_{eff}$ and $N(Q^2$) are given by,
\beqn
\sigma_{eff}(E=6\ GeV)=17\pm 2\ mb;\ \ \  N(E=6\ GeV)&=&(5.4\pm0.4)\zeta\;
\hskip .4in \chi^2=0.28 \nonumber \\
\sigma_{eff}(E=10\ GeV)=12\pm 2\ mb;\ \ \ N(E=10\ GeV)&=&(3.3\pm0.4)\zeta\;
\hskip .4in \chi^2=0.53\nonumber
\eeqn
where $\chi^2 = \Sigma[(y_i - d_i)/\Delta d_i]^2$, $d_i$ are the data points,
$\Delta d_i$ is the error in $d_i$  and $y_i$ are the theoretical values
calculated using the Monte-Carlo,
 and $\zeta = 5.2\times10^7$ mb GeV$^{18}$ is a constant
containing the  overall normalization of the  cross section.
The fit to the data clearly shows that $\sigma_{eff}$ decreases with
energy and is considerably lower than a typical inelastic
cross section of $36\ mb$.

Having fit the data's A dependence, the fit can be examined at fixed A to find
the survival factor as a  function of $Q^2$.  The BNL data was taken
only at two energies for a good range of A, so  we have $N(Q^2)$ and
$\sigma_{eff}(Q^2$) at
two points. However for the Aluminum target the data at intermediate
energies was also given.
We can make a function for $N(Q^2)$ which interpolates with $Q^2$
smoothly between the two endpoints where $Q^2$ was reported. It is given by:
\beq
N(Q^2) = {5.4 \zeta\over  (Q^24.8 GeV^2)^{0.86}}\ \ .
\eeq

Then inverting (1) we can determine the survival factor $P_{ppp}$ as
well as $\sigma_{eff}$ as a function of $Q^2$ by using the data for
Aluminum at intermediate energies. The
resulting best consistent fit for $P_{ppp}$ is shown in Fig. (2) and
 $\sigma_{eff}$($Q^2$) is shown  in Fig. (3). The
survival factor would be flat with $Q^2$ if one used  a traditional Glauber
model.
The fact that the survival probability rises with $Q^2$ is clear evidence for
color transparency.  It has gone unnoticed in studies involving
the transparency ratio,  because the assumed normalization in that case skews
the analysis.

\medskip

The effective attenuation cross section $\sigma_{eff}$
should be a universal quantity
which can be compared  from experiment to experiment.  The theoretical basis
for this is factorization between the hard  scattering and the nuclear
propagation factors. An
approximate fit to the attenuation cross section from the fit is
\beq
\sigma_{eff}(Q^2)= 40 mb (2.2 GeV^2/ Q^2)\pm 2 mb
\eeq
for 4.8 $GeV^2< Q^2 < 8.5 GeV^2$.
The decrease with $Q^2$ of the attenuation cross section coincides
with the rate predicted theoretically on the basis that the cross section goes
like the transverse  separations $b_T^2$ of the participating quarks, and
that the
region of important $b_T^2 = 1/Q^2$[4].   However, the scale in the
numerator of
2.2 GeV$^2$ was not predicted.  To put the scale in context,  it  says that the
perturbative QCD ideas are beginning to apply for Q$\geq$1.5 GeV.  This is
another
way  to present the conclusion that color transparency was actually observed in
the BNL experiment.

A final consistency check involves looking at the the normalization factor.
After taking out the  nuclear attenuation effects, according to the
perturbative treatment the $Q^2$ dependence of $N(Q^2)$ is  due to the hard
scattering process.  We have found that $N(Q^2)$ decreases relative
to $s^{-10}$,
meaning  that the hard scattering rate in the nuclear target is decreasing
faster than the naive quark counting  model prediction.  In perturbative QCD,
however, the quark-counting  prediction is modified,  due  to the running
coupling $\alpha_s(Q^2/\Lambda^2_{qcd})$ and scaling of distribution
amplitudes.  The perturbative
QCD prediction goes like $\alpha_s^{10}$ because there are five gluons in the
amplitude:
\beq d\sigma/dt_{pQCD} \cong
(\alpha_s(Q^2/\Lambda_{qcd}^2))^{10}s^{-10}f(t/s)\eeq
It becomes interesting to compare the form including powers of $\alpha_s$ with
the hard
scattering rate in  the nuclear target. (The $Q^2$ dependence of
$\alpha_s^{10}$ causes
serious disagreement with the data for  isolated pp scattering in free space
with $\Lambda_{qcd} \approx$ 100 MeV.)  The reason for consulting the
perturbative prediction
is to see whether the nuclear target has filtered the events down to something
like the shortest distance component.  Although (4) appears naive it is
 adequate because of  the usual ambiguity in the choice of scale.  We
must generate a range of reasonable theoretical  predictions by
choosing the $Q^2$ scale
of $\alpha_s$ in a typical range $(-t/2)\ <\ Q^2\ <\ (-1.5\ t)$.
  To improve on this
theoretically requires a calculation of next to leading logarithms.

\medskip

For comparison with the nuclear target we form the ratio of the global
$s^{-10}N(Q^2)$ to the pQCD  predictions (4) and plot the result
as solid lines in
Fig. (4). (The asymptotic prediction [5]
is also  illustrated as a dashed line to
show that it falls within the region of scale ambiguity of the pQCD
predictions.  For comparison we show s$^{-9.7}$, which is well outside the
range of the perturbative  predictions.)  The solid lines of
$s^{-10}N(Q^2)/(d\sigma/dt_{pQCD}(Q^2)$)
are rather flat, showing good  agreement of the BNL nuclear hard scattering
data
with QCD. For the Aluminum target we have  several data points at several $Q^2$
which also fall fairly well within the band of perturbative  predictions.
These are rather spectacular results, but we emphasize that they should be
viewed with  caution because the BNL experiment was a pioneering one.  If it is
confirmed by upcoming higher  precision data,  it will lead a great deal of
strength to the idea that QCD is cleaner after filtering in  a nuclear target.

\medskip

The procedure used here to
analyze the Brookhaven data can be applied
directly to quasi-elastic  electroproduction experiments such as the
SLAC-NE18.
For consistency, these experiments should see the same
attenuation cross sections at the
same $Q^2$. The free space scattering cross section in the case
of electroproduction is also
expected to receive significant contribution from large distance
components of the proton wave function.
Therefore, based on our experience with proton initiated
experiment, the hard scattering in this case should again be quite
different from free space scattering and should be closer to pQCD
predictions.

\medskip

In conclusion, we remark that the experimental determination of color
transparency seems to be  more subtle than was originally supposed.  The basic
issue is that exclusive hard scattering in free space gets significant
contribution from the soft components of proton wave function which are
absent in the nuclear medium because of filtering. The BNL data strongly
supports this idea since the hard
scattering rate in the nucleus turns out to be
 quite different from free space rate
and is very close to perturbative QCD predictions. The
BNL data also shows clear evidence for color transparency and
indicates an attenuation cross section that decreases with $Q^2$
at  the same rate as perturbatively predicted.
 We eagerly await new data, in  the hopes that it will confirm our
conclusion that color transparency was observed by  Carroll et.al.

{\bf Acknowledgements:}    This work has been supported in part by the DOE
Grant No.
DE-FG02-85-ER-40214.A008.

\bigskip

\noindent {\bf References}
\bigskip
\begin{enumerate}
\item S. J. Brodsky and A. H. Mueller, Phys. Lett. B {\bf 206}, 685 (1988), and
references therein.

\item  A. S. Carroll et al., Phys. Rev. Lett. {\bf 61}, 1698 (1988).

\item J. P. Ralston and B. Pire, Phys. Rev. Lett. {\bf 61}, 1823 (1988);
ibid {\bf 65}, 2343 (1990).

\item S. Nussinov, Phys. Rev. Lett {\bf 34}, 1286 (1975); F. E. Low , Phys.
Rev.
{\bf D12}, 163 (1975);  J.  Gunion and D. Soper, Phys. Rev. {\bf D15}, 2617
(1977).

\item S. J. Brodsky and G. P. Lepage, Phys. Rev. {\bf D22}, 2157 (1980).
\end{enumerate}
\end{document}